\documentclass[12pt]{iopart}

\usepackage{graphicx}

\begin{document}

\title[Conductance of carbon nanotubes decorated with gold clusters]
{Conductance of carbon nanotubes functionalized with gold clusters during CO adsorption}
\author{P Gurung and N Deo}
\address{Department of Physics and Astrophysics, University of Delhi, Delhi 110007, India}
\ead{ndeo@physics.du.ac.in}
\begin{abstract}
We investigate the time-dependent electronic transport in single-walled carbon nanotubes
(SWCNTs) functionalized with Au clusters on CO gas exposure. Using a tight-binding
Hamiltonian and the nonequilibrium Green's function (NEGF) formalism the time-dependent
zeroth and first order contributions to the current are calculated. The zeroth order
contribution is identified as the time-dependent Landauer formula in terms of the slow
time variable, whereas the first order contribution is found to be small. The Green's
function for the SWCNT is derived using the equation of motion and Dyson equation technique.
The conductance is explicitly evaluated by considering a form for the hopping integral which
accounts for the effect of dopants on the charge distribution of the carbon atoms and the
nearest-neighbor distance. The effect of dopants is also studied in terms of fluctuations
by calculating the autocorrelation function for the experimental and theoretical data. These
calculations allow direct comparison with the experiment and demonstrate how the presence of
dopants modifies the conductance of the SWCNT measured experimentally and provide the only
study of fluctuations in the sensor response in terms of the autocorrelation function.
\end{abstract}

\maketitle

\section{Introduction}
In the past years, carbon nanotubes (CNTs) \cite{1}, especially single-walled
carbon nanotubes (SWCNTs) have been actively studied as a chemical (gas) sensor
because of their unique nanostructure and electronic properties \cite{2}. Sensors
make significant impact in everyday life with applications ranging from health
to environment. Detection of hazardous gases using miniaturized sensing devices
with high sensitivity is an open challenge in chemical sensing applications for
environmental safety, including air-pollutants monitoring. 
As a result, research emphasis is on developing novel sensing materials and technologies.
The chemical sensing capabilities of CNTs based sensors can be amplified by their
functionalization. In recent years, carbon nanotubes decorated with metal nanoparticles (NPs)
have attracted a tremendous amount of interest and research activity in the applications
of CNTs as gas sensors. Previous studies have shown CNTs functionalized with metal
NPs exhibit unique sensitivity toward various gases \cite{3,4,5,6,7,8,9,10}. The
sensing capability of metal NP-decorated CNTs is based on the changes in their
electrical properties induced by gas molecules adsorbed on the NP surface
\cite{3,4,5,6,7,8,9,10}. The exceptional structural and electronic properties of
CNTs make them a potentially ideal material for the exploration of electronic
transport phenomena in low dimensional systems. In this context, nonequilibrium
Green's function (NEGF) formalism provides a powerful technique for the development
of quantitative models for quantum transport in such systems. The NEGF formalism
was developed by Kandanoff and Baym \cite{11}, and Keldysh \cite{12,12a} which provides a
microscopic theory to model quantum transport in mesoscopic semiconductor systems \cite{13,13a,13b,13c}.
Electronic transport in these systems is divided into stationary and time-dependent
phenomena.
The results of stationary transport using the NEGF formalism have been reported by many
authors \cite{14,14a,14b,15,16,17,18,19,20,21}.  In these studies, a key result was a
Landauer type formula \cite{22,22a} which gives a relationship between the conductance
of a mesoscopic sample, connected to the contacts by two leads, and its transmission
properties. This NEGF technique was also extended to study time-dependent transport in
mesoscopic systems \cite{23,24,25,26,27,27a}. A general formulation for time-dependent transport
through mesoscopic structures was introduced using time-dependent voltages in Refs. \cite{28,
29, 30}. Hernandez \emph{et al.} \cite{31} presented their study of time-dependent electronic
transport through a quantum dot using the NEGF formalism, whereas Kienle \emph{et al.}
\cite{32} studied the time-dependent quantum transport through a ballistic CNT transistor
in the presence of a time harmonic signal. The work presented in this paper is motivated
by these time-dependent results \cite{28,29,30,31,32} but distinguishes from these studies
in terms of time-dependence which arises due to interaction of CO gas molecules with
functionalized SWCNT not because of any externally applied time varying potentials.

We present a theory which explains the experimental result of the electronic transport
in SWCNTs decorated with Au clusters (Au-SWCNTs) in the presence of CO gas molecules \cite{10}.
To realize the application of SWCNTs as a gas sensor, a conceptual and quantitative understanding
of the underlying mechanism in experiments is necessary. There has not been much effort to model such devices
and this manuscript represents an important contribution to the field. The quantitative understanding
of the experiment involves building up a theoretical model of the Au functionalized SWCNT to study
electronic transport, and the development of a connection between the theoretical predictions and
experimental observations. The experiment is performed for many Au clusters (having different dimensions)
with 30 minutes of exposure to 2500 ppm CO gas at room temperature and a bias voltage of 0.5 V \cite{10}.
To model the electronic transport through the Au-SWCNT system on CO adsorption, the calculations are
performed for a defective (14,0) SWCNT decorated with an $\rm{Au_{20}}$ cluster. The model considered
in the calculations is based on the tight-binding Hamiltonian which describes electrons in SWCNT.
Analytical calculations of time-dependent electronic transport for the Au-SWCNT system are performed
using the NEGF formalism. In the system, there are two time scales: a slow time scale ($\bar t$) for
CO gas flow and a fast time scale ($t-t^\prime$) for electron transport inside the SWCNT. Hence, an
adiabatic expansion with respect to slow time variable and the Fourier transform with respect to fast
time variable are made in the calculations, leading to the zeroth and first order contributions to the
current. The zeroth order current is identified as the Landauer formula in terms of the slow time
variable and the first order contribution is found to be small. We derive an explicit formula for the
transmission function and calculate the conductance in terms of the time-dependent hopping integral
and on-site energy. The formula is then used to compare the theoretical results with the experiment
(i.e., conductance versus time plot) \cite{10}, by choosing a form for the hopping integral which
accounts for the effect of the Au and CO on the transport properties of the SWCNT.
The effect of the Au and CO molecules on electronic transport in SWCNT is also studied in terms of
fluctuations by calculating the autocorrelation function (ACF) for the theoretical and experimental
data.

\section{Tight-binding model and the nonequilibrium Green's function formalism}
The adsorption configuration of nine CO molecules attached to the corner (i) and edge
(ii, iii) sites of an $\rm Au_{20}$ cluster on a (14,0) SWCNT with a missing carbon atom
defect \cite{10} is shown in Fig. 1, where the C, Au and O atoms are shown in green, yellow
and red. This configuration demonstrates that the Au cluster affects only the first
nearest-neighbor carbon atoms of the SWCNT that causes changes in conductance.
\begin{figure}[h!]
\center
\includegraphics[width=9cm]{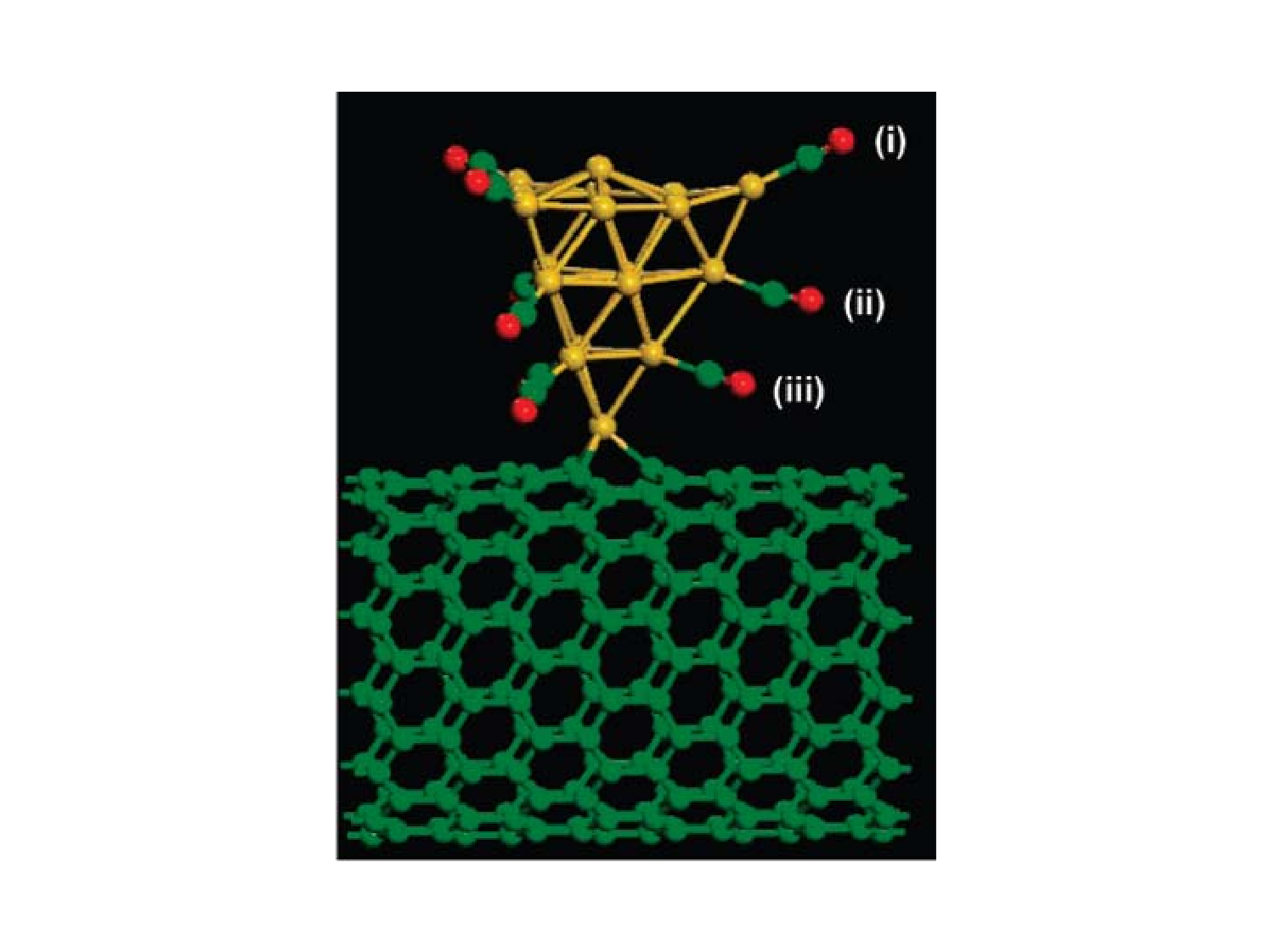}
\caption{(Color online) Adsorption configuration of nine CO molecules attached to the corner (i) and edge
(ii, iii) sites of an $\rm Au_{20}$ cluster on a defective (14,0) SWCNT, where the C, Au
and O atoms are shown in green, yellow and red color. Taken from Ref. \cite{10} with permission
from American Chemical Society.}
\end{figure}

To model the electronic transport in the system, a tight-binding model of the
$\rm {Au_{20}}$-SWCNT system (C) sandwiched between two semi-infinite (14,0)
left (L) and right (R) nanotube electrodes is presented. The chosen model
system is same as considered in the theoretical study of the experiment \cite{10}.
The model is shown in Fig. 2 where Au atoms are indicated by yellow solid circles,
C atoms are denoted by blue solid circles with an outline, and O atoms are indicated
by red empty circles. The model consists of only four nearest-neighbor carbon atoms
$\rm A_1$, $\rm B_1$, $\rm B_2$ and $\rm A_3$ of which only the the first nearest-neighbor
carbon atoms $\rm B_1$ and $\rm B_2$ are affected by the $\rm Au_{20}$ cluster
positioned at the missing carbon atom site $\rm A_2$. The second nearest-neighbor
carbon atoms $\rm A_1$ and $\rm A_3$ remain unaffected. If more carbon atoms that
are next nearest-neighbors to the $\rm Au_{20}$ cluster are considered in the model then there
will be no significant change in the result as only the first nearest-neighbor carbon
atoms contribute to the changes in the conductance. As a result, the effective length
of the nanotube involved in transport for sensing is a few nanometers.

The operational principle of the model is based on the changes in conductance when
the $\rm {Au_{20}}$-SWCNT system is exposed to CO gas for 30 minutes. On exposure to
CO gas, a CO molecule interacts with the $\rm {Au_{20}}$ cluster (positioned at site $\rm A_2$)
at an instant of time $\bar t_1$ which affects the hopping of electrons from the $\pi$
orbital of one carbon atom $\rm A_1$ to the neighboring carbon atom $\rm B_1$ with the
corresponding hopping integral $\gamma_{11}(\bar{t}_1)=\gamma_0$ (the hopping integral
of pristine SWCNT). Then, the electron hops from $\rm B_1$ to $\rm A_2$ with the hopping
integral $\gamma_{12}(\bar{t}_1)$, from $\rm A_2$ to $\rm B_2$ with $\gamma_{22}(\bar{t}_1)$,
and from $\rm B_2$ to $\rm A_3$ with $\gamma_{23}(\bar{t}_1)=\gamma_0$, Fig. 2. In a
similar way, the hopping integral changes when other CO molecules interact with the
$\rm {Au_{20}}$ cluster at the next instants of time $\bar t_2, \bar t_3$ and so on. 
This is how the time-dependence arises here without the application of a time-dependent bias.
The time-dependent hopping integral leads to time-dependent Hamiltonian and Green's functions
for the SWCNT. Thus, to study time-dependent transport through such a nano-hybrid model the
time-dependent NEGF formalism is well suited.

\begin{figure}[h!]
\center
\includegraphics[width=11cm]{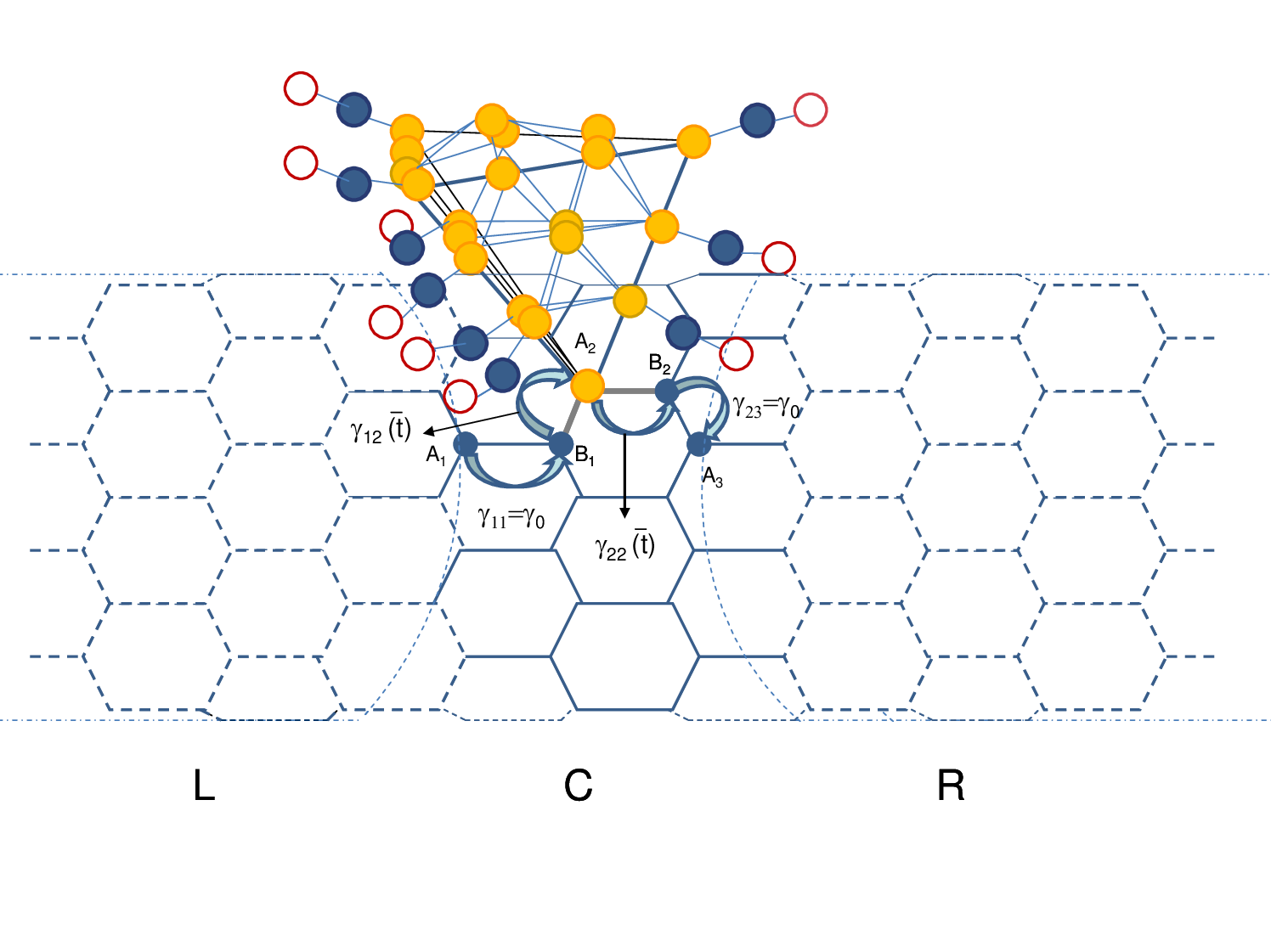}
\caption{(Color online) Schematic view of adsorption of CO molecules at the $\rm Au_{20}$ cluster surface on a
semiconducting (14,0) SWCNT with a missing carbon atom defect at site $\rm A_2$. Au atoms are
yellow solid circles, C atoms are blue solid circles with outline, and O atoms are red empty
circles. L, R indicate the left and right nanotube electrodes and C denotes the $\rm Au_{20}$-SWCNT
system. The arrow indicates the path of transmission.}
\end{figure}

\subsection{Model Hamiltonian}
To calculate the time-dependent conductance in the model system we use the nonequilibrium Green's
function (NEGF) formalism. The time-dependent electronic transport through mesoscopic and nano
scale systems has been addressed in the literature \cite{28,29,30,31,32} where the time-dependence
arises due to application of external time-dependent bias between the source and drain contacts, and
the coupling between the leads and the central region can be controlled by time-dependent
gate voltage. But this manuscript addresses the problem in which time-dependence arises due to interaction
of CO gas molecules with the Au-decorated SWCNT for a given gas exposure time. Therefore, the total
Hamiltonian of the model system corresponding to a semiconducting SWCNT (C) and the two left and right
nanotube electrodes L and R is expressed as
\begin{equation}
H_{\rm total}=H_{\rm SWCNT}+H_{\rm contact}+H_{\rm tunneling},
\end{equation}

where $H_{\rm SWCNT}$ is the Hamiltonian of the SWCNT and is given as
\begin{equation}
H_{\rm SWCNT}=\sum_m \varepsilon_m(t)d_m^\dagger d_m+\sum_{<mn>}
\gamma_{mn}(t)d_m^\dagger d_n,
\end{equation}
where $d_m^\dagger(d_m)$ creates (annihilates) an electron in state m of the
carbon lattice and $\varepsilon_m (t)$ is the on-site energy of the carbon atom,
whereas $\gamma_{mn}(t)$ is the nearest-neighbor hopping energy between the carbon
atoms as a function of time.
The effect of the Au and CO gas is included in the form of time-dependent
on-site energy and hopping integrals because as the time changes, the number of CO
molecules and hence their interaction with Au cluster changes, which results
in a change in on-site energy and hopping integral leading to a change in sensor response.

The contact Hamiltonian is written as
\begin{equation}
H_{\rm contact}=\sum_{k\alpha\epsilon {\rm L, R}} \varepsilon_{k\alpha}c^\dagger_{k\alpha} c_{k\alpha},
\end{equation}
where $c^\dagger_{k\alpha}$ and $c_{k\alpha}$ are the creation and annihilation operators
for electrons with momentum $k$ in either the L or R contact. It should be noted that there
is no time dependence in $H_{\rm contact}$, as in the experiment no external time-dependent
bias is applied between the L and R contacts, and electrons in the contacts are non-interacting.

The coupling between the contacts and the central SWCNT in the absence of a time-dependent
gate voltage is described by the tunneling Hamiltonian
\begin{equation}
H_{\rm tunneling}=H_T=
\sum_{n;k\alpha\epsilon {\rm L,R}} (V_{k\alpha, n}
c^\dagger_{k\alpha} d_n + H.c.).
\end{equation}

In the matrix form, the total Hamiltonian is expressed as
\begin{equation}
H({t}) = \left(\begin{array}{ccc}
H_{\rm L} & V_{\rm {LC}} & 0  \\
V^\dagger_{\rm {LC}} & H_{\rm C} ( t) & V_{\rm {CR}} \\
0 & V^\dagger_{\rm {CR}} & H_{\rm R} \\
\end{array} \right),
\end{equation}
where $H_{\{\rm L\rm R\}}$ are the left and right SWCNT contact Hamiltonian and $H_{\rm C}$
is the time-dependent tight-binding Hamiltonian for the central SWCNT. The matrices $V_{\rm {LC}}$
and $V_{\rm{CR}}$ are the tunneling (coupling) Hamiltonian between the two contacts and the central
SWCNT.

Since the $\rm Au_{20}$ cluster at site $\rm A_2$ affects only the first nearest-neighbor carbon atoms
$\rm B_1$ and $\rm B_2$ with the corresponding hopping integral $\gamma_{12}(t)$ and $\gamma_{22}(t)$,
as a function of time, the second nearest-neighbor carbon atoms $\rm A_1$ and $\rm A_3$ remain unchanged
with the hopping integrals $\gamma_0$. Therefore, the Hamiltonian of the Au-decorated SWCNT system with
CO adsorption is of the form
\begin{eqnarray}
H_{\rm C}(t)=
\left(
\begin{array}{ccccc}
\varepsilon_{\rm A_{1}} & {\gamma_{0}} &  &  &\\
{\gamma_{0}} & {\varepsilon_{\rm B_{1}}(t)} & {\gamma_{12}(t)} & &\\
& {\gamma_{12}(t)} & {\varepsilon_{\rm A_{2}}(t)} & {\gamma_{22}(t)} & \\
& & {\gamma_{22}(t)} & {\varepsilon_{\rm B_{2}}(t)} & {\gamma_{0}}\\
&&&{\gamma_{0}}&\varepsilon_{\rm A_3}
\end{array} \right).
\end{eqnarray}

\subsection{Expression for the current and equation of motion}

For this model system, an expression for the current flowing from the left contact to the central SWCNT
region is derived as \cite{29,30} 

\begin{eqnarray}
I_{\rm L}(t)
&=&\frac{2e}{\hbar}{\rm Re}\bigg\{\sum_{n;k\alpha\epsilon {\rm L}}V_{k\alpha,n}G^<_{n,k\alpha}(t,t)\bigg\},
\end{eqnarray}

where, the lesser Green's function is given as
\begin{eqnarray}
G^<_{n,k\alpha}(t, t^{\prime})&=&i\bigg<c^\dagger_{k\alpha}(t^{\prime})d_n(t)\bigg>.
\end{eqnarray}
%

\subsubsection{Equation of motion}

An expression for the $G^<_{n,k\alpha}(t,t)$ can be derived from the equation of motion for the
contact time-ordered Green's function $G_{n,k\alpha}(t,t^\prime)$ which is defined as \cite{31}
\begin{equation}
G_{n,k\alpha}(t,t^\prime)=\frac{-i}{\hbar}\bigg<{\rm T}[d_n(t)c^\dagger_{k\alpha}(t^\prime)]\bigg>,
\end{equation}

where T is the time-ordering operator defined as ${\rm T}[A(t)B(t^\prime)]$=$\theta(t-t^\prime)A(t)B(t^\prime)-\theta(t^\prime-t)B(t^\prime)
A(t)$ \cite{30}. The equation of motion for $G_{n,k\alpha}(t,t^\prime)$ 
is expressed as

\begin{eqnarray}
\fl
\bigg[-i\hbar\frac{\partial}{\partial t^\prime}-\varepsilon_{k\alpha}\bigg]G_{n,k\alpha}(t,t^\prime) 
&=&\sum_{m^\prime}V^*_{k\alpha, m^\prime}G_{nm^\prime}(t,t^\prime)=\sum_mV^*_{k\alpha,m}G_{nm}(t,t^\prime),
\end{eqnarray}

with $G_{nm}(t,t^\prime)$ the time-ordered Green's function of the central carbon nanotube defined as \cite{31}
\begin{equation}
G_{nm}(t,t^\prime)=\frac{-i}{\hbar}<{\rm T}[d_n(t)d_m^\dagger(t^\prime)]>,
\end{equation}
which satisfies the equation of motion
\begin{eqnarray}
\bigg[i\hbar\frac{\partial}{\partial t}-\varepsilon_n(t)\bigg]G_{nm}(t,t^\prime) 
&=&\delta (t-t^\prime)\delta_{nm}+\sum_{<n n^\prime>}\gamma_{n n^\prime}(t)G_{n^\prime m}(t,t^\prime) \nonumber\\
&&+\sum_{k^\prime\alpha^\prime}V^*_{k^\prime\alpha^\prime, n}G_{k^\prime\alpha^\prime,m}(t,t^\prime).
\end{eqnarray}

Equation (12) can be further written as

\begin{eqnarray}
\fl
\bigg[i\hbar\frac{\partial}{\partial t}\delta_{nn^\prime}-\varepsilon_n(t)\delta_{nn^\prime}-\sum_{<n n^\prime>}\gamma_{n n^\prime}(t)\bigg]G_{n^\prime m}(t,t^\prime)&=& \delta (t-t^\prime)\delta_{nm}\nonumber\\
&&+ \sum_{k^\prime\alpha^\prime}V^*_{k^\prime\alpha^\prime, n}G_{k^\prime\alpha^\prime,m}(t,t^\prime).
\end{eqnarray}
The time-ordered Green's function, Eq(10), is further written as
\begin{equation}
G_{n,k\alpha}(t,t^\prime)=\sum_m\int dt_1 G_{nm}(t,t_1)V^*_{k\alpha,m}g_{k\alpha}(t_1-t^\prime),
\end{equation}

where $g_{k\alpha}(t_1-t^\prime)$ is the contact Green's functions operator for the uncoupled system \cite{30}.
Equation (14) gives rise to the lesser Green's function \cite{30}
\begin{eqnarray}
\fl
G^<_{n,k\alpha}(t,t)
&=&\sum_m\int dt^{\prime} V^*_{k\alpha,m} e^{-i\varepsilon_{k\alpha}(t^\prime-t)}\bigg\{G^r_{nm}(t,t^\prime)if(\varepsilon^0_{k\alpha})+i\theta(t- t^\prime)G^<_{nm}(t,t^\prime)\bigg\},
\end{eqnarray}
where the retarded and lesser Green's functions of the central carbon nanotube are defined as
\begin{eqnarray}
G^r_{nm}(t, t^\prime)&=&\frac{-i}{\hbar}\theta(t-t^\prime)<\{d_n(t),d^\dagger_m(t^\prime)\}> ~{\rm and}~~~~\nonumber\\
G^<_{nm}(t,t^\prime)&=&\frac{i}{\hbar}<d^\dagger_m(t^\prime)d_n(t)>.
\end{eqnarray}

Putting Eq. (15) in Eq. (7) and converting the sum over the momentum states $k$ in the contacts into an
integral over energies, Eq. (7) is expressed as

%

\begin{eqnarray}
\fl
I_{\rm L}(t)&=&\frac{2e}{\hbar}\int dt^\prime\int \frac {d\varepsilon}{2\pi}{\rm Re}\bigg\{e^{-i\varepsilon_{k\alpha}(t^\prime-t)}[\Gamma_{\rm L}(\varepsilon)]_{mn}\bigg[G^r_{nm}(t,t^\prime)i f_{\rm L}(\varepsilon) 
+i\theta(t-t^\prime)G^<_{nm}(t,t^\prime)\bigg]\bigg\},
\end{eqnarray}

where $[\Gamma_{\rm L}(\varepsilon)]_{mn}$ is the time-independent coupling between the left contact and
the central carbon nanotube given as
\begin{equation}
[\Gamma_{\rm L}(\varepsilon)]_{mn}=2\pi \sum_{\alpha\epsilon {\rm L}} \rho_\alpha(\varepsilon) V_{\alpha,n}(\varepsilon)V^*_{\alpha,m}(\varepsilon),
\end{equation}
 with $\rho_\alpha(\varepsilon)$ the density of states.

Splitting the limit over the time integral 
the expression for the current becomes
\begin{eqnarray}
\fl
I_{\rm L}(t)=&&\frac{-2e}{\hbar}\int^t_{-\infty} dt^\prime\int\frac{d\varepsilon}{2\pi} {\rm Im } \bigg\{e^{-i\varepsilon_{k\alpha}(t^\prime-t)}[\Gamma_{\rm L}(\varepsilon)]_{mn} 
\bigg[f_{\rm L}(\varepsilon)G^r_{nm}(t,t^\prime)+ G^<_{nm}(t,t^\prime)\bigg]\bigg\}.
\end{eqnarray}

In matrix form, the current is expressed as
\begin{eqnarray}
\fl
I_{\rm L}(t)=&&\frac{-2e}{\hbar}\int^t_{-\infty} dt^\prime\int\frac{d\varepsilon}{2\pi}{\rm Im Tr} \bigg\{e^{-i\varepsilon_{k\alpha}(t^\prime-t)}{\Gamma}_{\rm L}(\varepsilon) \bigg[f_{\rm L}(\varepsilon){G}^r_{\rm C}(t,t^\prime)+{G}^<_{\rm C}(t,t^\prime)\bigg]\bigg\},
\end{eqnarray}
where $G_{\rm C}$ denotes the Green's function of the carbon nanotube in the presence of the
coupling with the contacts.

Hence, the current flowing from the left/right contact to the central carbon nanotube is written as
\begin{eqnarray}
\fl
I_{\rm L/R}(t)&=&\frac{-2e}{\hbar}\int^t_{-\infty} dt^\prime\int\frac{d\varepsilon}{2\pi}{\rm Im Tr} \bigg\{e^{-i\varepsilon_{k\alpha}(t^\prime-t)}{\Gamma}_{\rm L/R}(\varepsilon)\bigg[f_{\rm L/R}(\varepsilon){G}^r_{\rm C}(t,t^\prime)+{G}^<_{\rm C}(t,t^\prime)\bigg]\bigg\}.
\end{eqnarray}

\subsection{Adiabatic approximation}
In the adiabatic approximation, the time scale over which the system parameters change is large
compared to the life time of an electron in the system (CNT) \cite{31}. In the problem of electronic
transport through the SWCNT-based sensor the experimental time scale of the gas interaction with the
Au-SWCNT is much longer ($\sim \rm seconds$) than the time scale of electron transport inside the SWCNT
($\sim 10^{-15}~ \rm s$). Therefore the adiabatic approximation is applied to study the electronic
transport in the gas sensor.

\subsubsection{Adiabatic expansion for the Green's functions}
The best way to separate slow and fast time scales is to re-parameterize the Green's functions.
In other words the time variables of the Green's functions are replaced by a fast time difference
$\delta t=t-t^{\prime}$ and a slow mean time $\bar t=\frac{t+t^{\prime}}{2}$ \cite {31} as
\begin{equation}
G(t, t^\prime)=G\bigg(t-t^\prime, \frac{t+t^\prime}{2}\bigg).
\end{equation}

The adiabatic approximation is applied to lowest order by expanding the Green's functions about
the slow time variable up to linear order in the fast time variables \cite{31}
\begin{eqnarray}
G(t-t^\prime,\bar t)&=&G(t-t^\prime, \bar t)|_{\bar t =t}+\Big(\frac{t^\prime-t}{2}\Big)\frac{\partial G}{\partial \bar t}(t-t^\prime,\bar t)|_{\bar t=t}\nonumber\\
&=& G(t-t^\prime,t)+ \Bigg(\frac{t^\prime-t}{2}\Bigg)\frac{\partial G}{\partial \bar t}(t-t^\prime,\bar t)|_{\bar t=t},
\end{eqnarray}
which can be further written as
\begin{equation}
G(t-t^\prime,\bar t)= G^{(0)}(t-t^\prime, \bar t)+G^{(1)}(t-t^\prime, \bar t),
\end{equation}

where $G^{(0)}(t-t^\prime, \bar t)$ and $G^{(1)}(t-t^\prime, \bar t)$ are the
zeroth and first order Green's functions which lead to the zeroth and first order
contributions to the current.

Applying adiabatic expansion on $G(t,t^\prime)$ in Eq. (21) the current becomes
\begin{eqnarray}
\fl
I_{\rm L/R}(t-t^\prime,\bar t)&=&\frac{-2e}{\hbar}\int^t_{-\infty} dt^\prime\int\frac{d\varepsilon}{2\pi}{\rm Im Tr} \bigg\{e^{-i\varepsilon_{k\alpha}(t^\prime-t)}{\Gamma}_{\rm L/R}(\varepsilon)\bigg[f_{\rm L/R}(\varepsilon)G^r_{\rm C}(t-t^\prime, \bar t)\nonumber\\
&&+G^<_{\rm C}(t-t^\prime,\bar t)\bigg]\bigg\}\nonumber\\
&=&\frac{-2e}{\hbar}\int^t_{-\infty} dt^\prime\int\frac{d\varepsilon}{2\pi}{\rm Im Tr} \bigg\{e^{-i\varepsilon_{k\alpha}(t^\prime-t)}{\Gamma}_{\rm L/R}(\varepsilon)\bigg[f_{\rm L/R}(\varepsilon)\bigg({G}^{(0)r}_{\rm C}(t-t^\prime, \bar t)\nonumber\\
&&+{G}^{(1)r}_{\rm C}(t-t^\prime, \bar t)\bigg)+\bigg(G^{(0)<}_{\rm C}(t-t^\prime,\bar t)+G^{(1)<}_{\rm C}(t-t^\prime,\bar t)\bigg)\bigg]\bigg\}.
\end{eqnarray}

Consider only the zeroth order contribution to the current
\begin{eqnarray}
\fl
I^{(0)}_{\rm L/R}(t-t^\prime,\bar t)&=&\frac{-2e}{\hbar}\int^t_{-\infty} dt^{\prime}\int\frac{d\varepsilon}{2\pi}{\rm Im Tr} \bigg\{e^{-i\varepsilon_{k\alpha}(t^\prime-t)}{\Gamma}_{\rm L/R}(\varepsilon)\bigg[G^{(0)<}_{\rm C}
(t-t^\prime,\bar t)\nonumber\\
&&+ f_{\rm L/R}(\varepsilon){G}^{(0)r}_{\rm C}(t-t^\prime, \bar t)\bigg]\bigg\}\nonumber\\
&=&I^{(0)}_1(t-t^\prime,\bar t)+I^{(0)}_2(t-t^\prime,\bar t).
\end{eqnarray}

Solving Eq. (26) by taking the Fourier transform and using Eq. (16) we obtain
\begin{eqnarray}
\fl
I^{(0)}_{\rm L/R}(\bar t)&=&\frac{ie}{\hbar}\int\frac{d\varepsilon}{2\pi}{\rm Tr}\bigg({\Gamma}_{\rm L/R}(\varepsilon)\bigg\{G^{(0)<}
_{\rm C}(\varepsilon,\bar t)+f_{\rm L/R}(\varepsilon)\bigg[G^{(0)r}_{\rm C}(\varepsilon, \bar t)-G^{(0)a}_{\rm C}(\varepsilon, \bar t)\bigg]\bigg\}\bigg).
\end{eqnarray}

\subsection{Green's function for the nanotube}

In Eq. (13), substituting $G_{k^\prime\alpha^\prime,m}(t,t^\prime)=g_{k^\prime\alpha^\prime}(t-t^\prime)\sum_{m^\prime} V_{k^\prime\alpha^\prime,m^\prime}G_{m^\prime m}(t,t^\prime)$ 
one gets

\begin{eqnarray}
\fl
\bigg[i\hbar\frac{\partial}{\partial t}\delta_{nn^\prime}-\varepsilon_n(t)\delta_{nn^\prime}-\sum_{<n n^\prime>}\gamma_{n n^\prime}(t)\bigg]G_{n^\prime m}(t,t^\prime)&=& \delta (t-t^\prime)\delta_{nm}+ \sum_{k^\prime\alpha^\prime}V^*_{k^\prime\alpha^\prime, n} g_{k^\prime\alpha^\prime}(t-t^\prime)\nonumber\\
&&\times\sum_{m^\prime} V_{k^\prime\alpha^\prime, m^\prime}G_{m^\prime m}(t,t^\prime)\nonumber\\
&=& \delta (t-t^\prime)\delta_{nm}+ \sum_{m^\prime}\Sigma_{nm^\prime}(t-t^\prime)\nonumber\\
&&\times G_{m^\prime m}(t,t^\prime),
\end{eqnarray}
where
\begin{equation}
\Sigma_{nm^\prime}(t-t^\prime)=\sum_{k^\prime\alpha^\prime\epsilon {\rm L,R}}V^*_{k^\prime\alpha^\prime,n}g_{k^\prime\alpha^\prime}(t-t^\prime)V_{k^\prime\alpha^\prime,m^\prime},
\end{equation}

is the time-independent self-energy \cite{30}.

After replacing $m^\prime$ with $n^\prime$ Eq. (28) can be further written as
\begin{eqnarray}
\fl
\bigg[i\hbar\frac{\partial}{\partial t}\delta_{nn^\prime}-\varepsilon_n(t)\delta_{nn^\prime}-\sum_{<n n^\prime>}\gamma_{n n^\prime}(t)\bigg]
G_{n^\prime m}(t,t^\prime)=\delta (t-t^\prime)\delta_{nm}+ \Sigma_{nn^\prime}G_{n^\prime m}(t,t^\prime).
\end{eqnarray}

Equation (30) can be rewritten in full matrix form as
\begin{equation}
\fl
\bigg[i\hbar\frac{\partial}{\partial t}-H_{\rm C}(t)-\Sigma_{\rm L}(t-t^\prime)-\Sigma_{\rm R}(t-t^\prime)\bigg]G_{\rm C}(t,t^\prime)=\delta(t-t^\prime)\delta_{nm}.
\end{equation}

Applying adiabatic expansion on $G_{\rm C}(t,t^\prime)$ one obtains 
\begin{equation}
\fl
\bigg[i\hbar\frac{\partial}{\partial t}-H_{\rm C}(t)-\Sigma_{\rm L}(t-t^\prime)-\Sigma_{\rm R}(t-t^\prime)\bigg]\bigg(G^{(0)}_{\rm C}(t-t^\prime,\bar t)+G^{(1)}_{\rm C}(t-t^\prime,\bar t)\bigg)=\delta(t-t^\prime)\delta_{nm}.
\end{equation}

Expanding $H_{\rm C}(t)$ around $\frac{t+t^\prime}{2}=\bar t$ \cite{31} gives rise to
\begin{equation}
H_{\rm C}(t)=H_{\rm C}(\bar t)+\frac {\partial H_{\rm C}(\bar t)}{\partial \bar t}(t-\bar t)=H^{(0)}_{\rm C}(\bar t)+H^{(1)}_{\rm C}(\bar t).
\end{equation}

Substituting Eq. (33) in Eq. (32) to get
\begin{equation}
\fl
\bigg[i\hbar\frac{\partial}{\partial t}-H^{(0)}_{\rm C}(\bar t)-H^{(1)}_{\rm C}(\bar t)-\Sigma_{\rm L}(t-t^\prime)-\Sigma_{\rm R}(t-t^\prime)\bigg]\bigg(G^{(0)}_{\rm C}(t-t^\prime,\bar t)+G^{(1)}_{\rm C}(t-t^\prime,\bar t)\bigg)=\delta(t-t^\prime)\delta_{nm}.
\end{equation}

Consider only the zeroth order contribution gives
\begin{eqnarray}
\fl
\bigg[i\hbar\frac{\partial}{\partial t}-H^{(0)}_{\rm C}(\bar t)-\Sigma_{\rm L}(t-t^\prime)-\Sigma_{\rm R}(t-t^\prime)\bigg]
&&\times G^{(0)}_{\rm C}(t-t^\prime,\bar t)=\delta(t-t^\prime)\delta_{nm}.
\end{eqnarray}

Taking the Fourier transform with respect to the fast time variable $(t-t^\prime)$ we get
\begin{equation}
\bigg[\varepsilon-H^{(0)}_{\rm C}(\bar t)-\Sigma_{\rm L}(\varepsilon)-\Sigma_{\rm R}(\varepsilon)\bigg]G^{(0)}_{\rm C}(\varepsilon,\bar t)=1.
\end{equation}

This leads to the zeroth order Green's function for the nanotube in the presence of
the coupling with the contacts
\begin{equation}
G_{\rm C}^{(0)}(\varepsilon,\bar t)=\frac{1}{\varepsilon-H^{(0)}_{\rm C}(\bar t)-\Sigma_{\rm L}(\varepsilon)-\Sigma_{\rm R}(\varepsilon)}.
\end{equation}

\subsection{Dyson equation}

Equation (30) in full matrix form is expressed as

\begin{equation}
\fl
\bigg[i\hbar\frac{\partial}{\partial t}-\varepsilon_n(t)-\sum_{<nn^\prime>}\gamma_{nn^\prime}(t)\bigg]G_{\rm C}(t,t^\prime)=\delta (t-t^\prime)\delta_{nm}+ \Sigma(t-t^\prime)G_{\rm C}(t,t^\prime).
\end{equation}


Define two auxillary time-ordered Green's functions $\rm g$ and $\rm \bar g$ that satisfy the equation of motions
\begin{equation}
\bigg[i\hbar\frac{\partial}{\partial t}-\varepsilon_n(t)\bigg]{\rm g}(t,t^{\prime})=\delta(t-t^{\prime}),
\end{equation}

and
\begin{equation}
\bigg[i\hbar\frac{\partial}{\partial t}-\varepsilon_n(t)-\sum_{<nn^\prime>}\gamma_{nn^\prime}(t)\bigg]{\rm \bar g}(t,t^\prime)=\delta(t-t^\prime), 
\end{equation}

which is rearranged to
\begin{equation}
\bigg[i\hbar\frac{\partial}{\partial t}-\varepsilon_n(t)\bigg]{\rm \bar g}(t,t^\prime)=\delta(t-t^\prime)+\sum \gamma _{nn^\prime}(t){\rm \bar g}(t,t^\prime).
\end{equation}

Using Eq. (39), Eq. (41) can be further written as
\begin{equation}
{\rm \bar g}(t,t^\prime)={\rm g}(t,t^\prime)\bigg[\delta(t-t^\prime)+\sum \gamma _{nn^\prime}(t){\rm \bar g}(t,t^\prime)\bigg].
\end{equation}

From Eqs. (38) and (40)
\begin{equation}
G_{\rm C}(t,t^\prime)={\rm \bar g}(t,t^\prime)+{\rm \bar g}(t,t^\prime)\Sigma(t-t^\prime)G_{\rm C}(t,t^\prime),
\end{equation}

which is the Dyson equation for the system of gas sensor.

\section{Results and Discussions}

\subsection{Calculation of the zeroth order time-dependent Green's function}
Expanding the Hamiltonian $H_{\rm C}(t)$ in Eq. (6) around $\bar t$ using Eq. (33), and taking only
the zeroth order contribution, the zeroth order Hamiltonian with respect to the slow time variable,
$H^{(0)}_{\rm C}(\bar t)$ is obtained. Using Eq. (37) and the Hamiltonian $H^{(0)}_{\rm C}(\bar t)$
the zeroth order time-dependent retarded Green's function $G^{(0)r}_{\rm C}$ for the SWCNT
is explicitly written in a $5 \times 5$ matrix form as

\begin{equation}
\fl
G^{(0) r}_{\rm C}(\varepsilon,\bar{t})= \\
\left(
\begin{array}{ccccc}
\varepsilon-\varepsilon_{A_{1}}-\Sigma^r_{\rm L}(\varepsilon) & {\gamma_{0}} & 0 & 0 & 0 \\
{\gamma_{0}} & \varepsilon-{\varepsilon_{B_{1}}(\bar{t})} & {\gamma_{12}(\bar{t})} & 0 & 0 \\
0 & {\gamma_{12}(\bar{t})} & \varepsilon-{\varepsilon_{A_{2}}(\bar{t})} & {\gamma_{22}(\bar{t})} & 0\\
0 & 0 & {\gamma_{22}(\bar{t})} & \varepsilon-{\varepsilon_{B_{2}}(\bar{t})} & {\gamma_{0}} \\
0 & 0 & 0 & {\gamma_{0}} & \varepsilon-{\varepsilon_{A_{3}}}-\Sigma^r_{\rm R}(\varepsilon)\\
\end{array} \right)^{-1}.
\end{equation}

\subsection{Zeroth order time-dependent Landauer formula}

Using Eq. (27) and writing the total current as $I^{(0)}(\bar t)= x I_{\rm L}^{(0)}-(1-x)I_{\rm R}^{(0)}$,
and assuming the left and right coupling functions are proportional to each other $({\Gamma}_{\rm L}(\varepsilon)
=\lambda{\Gamma}_{\rm R}(\varepsilon))$, where $x=\frac{1}{1+\lambda}$ is the arbitrary parameter with $\lambda$
the constant of proportionality \cite{30}, a simple expression for the total current through the SWCNT is derived as

%
\begin{equation}
\fl
I^{(0)}(\bar t )=\frac{ie}{\hbar}\int\frac{d\varepsilon}{2\pi} {\rm Tr}\bigg\{\frac{\Gamma_{\rm L}(\varepsilon)
\Gamma_{\rm R}(\varepsilon)}{\Gamma_{\rm L}(\varepsilon)+\Gamma_{\rm R}(\varepsilon)}\bigg({G}^{(0)r}_{\rm C}(\varepsilon, \bar t)
-{G}^{(0)a}_{\rm C}(\varepsilon, \bar t))\bigg)\bigg\}[f_{\rm L}(\varepsilon)-f_{\rm R}(\varepsilon)].
\end{equation}

Applying the adiabatic expansion and taking the Fourier transform of the Dyson equation (Eq. (43)) for the retarded
Green's function, one gets
\begin{equation}
G^r_{\rm C}(\varepsilon,\bar t)={\rm \bar g}^r(\varepsilon,\bar t)+{\rm \bar g}^r(\varepsilon,\bar t)\Sigma^r(\varepsilon)G^r_{\rm C}(\varepsilon,\bar t),
\end{equation}

where $\Sigma^r(\varepsilon)$ the self-energy can be expressed as \cite{17}
\begin{equation}
\Sigma^r(\varepsilon)=\frac{G^{r}_{\rm C}(\varepsilon,\bar t)-{\rm \bar g}^r(\varepsilon,\bar t)}{{\bar g}^r(\varepsilon,\bar t)G^{r}_{\rm C}(\varepsilon,\bar t)}={\rm \bar g}^r(\varepsilon,\bar t)^{-1}-G^{r}_{\rm C}(\varepsilon,\bar t)^{-1}.
\end{equation}

In this model system, the effect of CO interaction with the Au-SWCNT is incorporated in $H^{(0)}_{\rm C}(\bar t)$
not in the self-energy $(\Sigma)$ which is time-independent and includes only the tunneling contributions with
no interaction. The retarded and advanced self-energies are defined as
\begin{equation}
\Sigma^{r,a}(\varepsilon)=\sum_{k\alpha\epsilon{\rm L, R}}|V_{k\alpha}|^2g^{r,a}_{k\alpha}(\varepsilon)=\Lambda(\varepsilon)\mp i\Gamma(\varepsilon)/2,
\end{equation}
where $\Lambda(\varepsilon)$ and $\Gamma(\varepsilon)/2$ are the real and imaginary parts of the self-energy
with $\Lambda(\varepsilon)=\Lambda_{\rm L}(\varepsilon)+\Lambda_{\rm R}(\varepsilon)$ and $\Gamma(\varepsilon)=\Gamma_{\rm L}(\varepsilon)
+\Gamma_{\rm R}(\varepsilon)$ \cite{30}.

This leads to the self-energy difference as \cite{17}
\begin{equation}
\Sigma(\varepsilon)=\Sigma^r(\varepsilon)-\Sigma^a(\varepsilon)=-i\Gamma(\varepsilon).
\end{equation}

Using Eq. (48) the zeroth order retarded and advanced Green's functions for the SWCNT from Eq. (37) are given as
\begin{equation}
G^{(0)r,a}_{\rm C}(\varepsilon,\bar t)=[\varepsilon-H^{(0)}_{\rm C}(\bar t)-\Lambda(\varepsilon)\pm i\Gamma(\varepsilon)/2]^{-1}.
\end{equation}

Equation (50) gives rise to

\begin{equation}
G^{(0)r}_{\rm C}(\varepsilon,\bar t)G^{(0)a}_{\rm C}(\varepsilon,\bar t)=\frac{1}{(\varepsilon-H^{(0)}_{\rm C}(\bar t)-\Lambda(\varepsilon))^2+(\Gamma(\varepsilon)/2)^2},
\end{equation}

and

\begin{equation}
G^{(0)r}_{\rm C}(\varepsilon,\bar t)-G^{(0)a}_{\rm C}(\varepsilon,\bar t)=\frac{-i\Gamma(\varepsilon)}{[\varepsilon-H^{(0)}_{\rm C}(\bar t)-\Lambda(\varepsilon)]^2+[\Gamma(\varepsilon)/2]^2}.
\end{equation}

From Eqs. (51) and (52)
\begin{equation}
G^{(0)r}_{\rm C}(\varepsilon,\bar t)-G^{(0)a}_{\rm C}(\varepsilon,\bar t)=G^{(0)r}_{\rm C}(\varepsilon,\bar t)(-i\Gamma(\varepsilon))G^{(0)a}_{\rm C}(\varepsilon,\bar t)
\end{equation}

Using Eq. (49), Eq. (53) becomes
\begin{equation}
G^{(0)r}_{\rm C}(\varepsilon,\bar t)-G^{(0)a}_{\rm C}(\varepsilon,\bar t)
=G^{(0)r}_{\rm C}(\varepsilon,\bar t)\Sigma(\varepsilon) G^{(0)a}_{\rm C}(\varepsilon,\bar t).
\end{equation}

Substituting Eq. (54) in Eq. (45)

\begin{eqnarray*}
\fl
I^{(0)}(\bar t)&=&\frac{e}{\hbar}\int\frac{d\varepsilon}{2\pi} {\rm Tr} \bigg\{\frac{i\Gamma_{\rm L}(\varepsilon)\Gamma_{\rm R}(\varepsilon)}
{\Gamma_{\rm L}(\varepsilon)+\Gamma_{\rm R}(\varepsilon)}{G}^{(0)r}_{\rm C}(\varepsilon, \bar t)\Sigma(\varepsilon)
{G}^{(0)a}_{\rm C}(\varepsilon, \bar t)\bigg\}[f_{\rm L}(\varepsilon)-f_{\rm R}(\varepsilon)].
\end{eqnarray*}

In this equation, the quantity $-i(\Gamma_{\rm L}(\varepsilon)+\Gamma_{\rm R}(\varepsilon))=-i\Gamma(\varepsilon)$
is just the self-energy $\Sigma(\varepsilon)$ (Eq. (49)). Hence the current becomes

\begin{eqnarray}
\fl
I^{(0)}(\bar t)&=& \frac{e}{\hbar}\int\frac{d\varepsilon}{2\pi} {\rm Tr}\bigg\{\Gamma_{\rm L}(\varepsilon)
{G}^{(0)r}_{\rm C}(\varepsilon, \bar t)\Gamma_{\rm R}(\varepsilon)
{G}^{(0)a}_{\rm C}(\varepsilon, \bar t)\Sigma(\varepsilon)
\Sigma^{-1}(\varepsilon)\bigg\}[f_{\rm L}(\varepsilon)-f_{\rm R}(\varepsilon)],
\end{eqnarray}

which leads to the current
\begin{equation}
I^{(0)}(\bar t)=\frac{e}{\hbar}\int\frac{d\varepsilon}{2\pi} {\rm Tr} \bigg\{\Gamma_{\rm L}(\varepsilon)
{G}_{\rm C}^{(0)r}(\varepsilon, \bar t)\Gamma_{\rm R}(\varepsilon){G}_{\rm C}^{(0)a}(\varepsilon, \bar t)
\bigg\}[f_{\rm L}(\varepsilon)-f_{\rm R}(\varepsilon)].
\end{equation}

This is the time-dependent Landauer formula which is derived in terms of the slow time variable $\bar t$,
where the transmission function of the system is identified as
\begin{equation}
T(\varepsilon,\bar{t})={\rm Tr}[{\Gamma_{\rm L}(\varepsilon)}{{G}_{\rm C}^{(0){r}}(\varepsilon,\bar{t})}{\Gamma_{\rm R}
(\varepsilon)}{{G}_{\rm C}}^{(0){a}}(\varepsilon,\bar{t})],
\end{equation}

with $f_{\{\rm L, R\}}(\varepsilon)$ are the Fermi distribution functions in the left and right electrodes,
and $\Gamma_{\{\rm L, R\}}=i[\Sigma^r_{\{\rm L, R\}}-\Sigma^a_{\{\rm L, R\}}]$ describe the contact-SWCNT coupling.

If only the first element of $\Gamma_{\rm L}(\varepsilon) (\Sigma_{\rm L}(\varepsilon))$ matrix and the last
element of $\Gamma_{\rm R}(\varepsilon)(\Sigma_{\rm R}(\varepsilon))$ matrix are considered, the transmission
function depends only on the off diagonal elements of the $G^{(0)}_{\rm C}(\varepsilon,\bar t)$
matrix and is expressed as
\begin{equation}
T(\varepsilon,\bar{t})=\Gamma_{\rm L,11}(\varepsilon)G^{(0)r}_{\rm C15}(\varepsilon,\bar{t})
\Gamma_{\rm R,55}(\varepsilon)G^{(0){r}*}_{\rm C15}(\varepsilon,\bar{t}),
\end{equation}

where $G^{(0)r*}=G^{(0)a}$.

To calculate the conductance explicitly, an expression for the transmission function is derived in the linear
response \cite{33} regime as the experiment is performed with low bias. In linear response, the expression for
the current (Eq.(56)) becomes $I^{(0)}(\bar{t})\approx\frac{e}{\hbar}\int{\frac{d{\varepsilon}}{2\pi}}T({\varepsilon},\bar{t})\delta
[f_{\rm L}(\varepsilon)-f_{\rm R}(\varepsilon)]$ which leads to the conductance given by

\begin{eqnarray}
{\cal G}^{(0)}(\bar t)&=&\frac{I^{(0)}(\bar{t})}{(\mu_{\rm L}-\mu_{\rm R})/e}=\frac{e^2}
{\hbar}\int{\frac{d{\varepsilon}}{2\pi}} T(\varepsilon,\bar{t})(-\frac{\partial f_0}{\partial\varepsilon}) 
=\frac{e^2}{\hbar} T(\varepsilon_f,\bar{t}),
\end{eqnarray}

as $\delta[f_{\rm L}(\varepsilon)-f_{\rm R}(\varepsilon)]=(\mu_{\rm L}-\mu_{\rm R})
(-\frac{\partial f_0}{\partial \varepsilon})$ and $(-\frac{\partial f_0}{\partial\varepsilon})=\delta(\varepsilon_f-\varepsilon)$
where $\mu_{\{\rm L, R\}}$ are the chemical potentials of the left and right contacts.

Transport is often dominated by states close to the Fermi level, and $\Gamma(\varepsilon)$
and $\Lambda(\varepsilon)$ the imaginary and real parts of the self-energy are generally
slowly varying functions of energy, therefore the wide-band limit is considered in
which the real part ($\Lambda$) of the self-energy is neglected and the imaginary
part ($\Gamma$) is considered to be energy independent. This approximation has the
advantage of providing explicit analytic results \cite{29,31}. Hence, $\Lambda$ is
neglected and $\Gamma$ is considered energy independent from now onwards.
Using the relationship $\Gamma_{\{\rm L, R\}}=i[\Sigma^r_{\{\rm L, R\}}-\Sigma^a{_{\{\rm L, R\}}}]$
the $\Gamma$ matrix elements are $\Gamma_{\rm L,11}=-2 {\rm Im}(\Sigma^r_{\rm L,11})=-2
{\rm Im}\Sigma^r_{\rm L}$ and $\Gamma_{\rm R,55}=-2{\rm Im}(\Sigma^r_{\rm R,55})=-2 {\rm Im}\Sigma^r_{\rm R}$.
Making use of these matrix elements, Eqs. (44) and (58), an expression for the transmission
function for the $\rm {Au_{20}}$-SWCNT system with CO gas in the linear response regime is given by

\begin{equation}
\label{eq.7}
T(\varepsilon_f,\bar{t})=\frac{4~{\rm Im}{\Sigma^r_{\rm L}}~{\rm Im}{\Sigma^r_{\rm R}
}~{\gamma^2_0}{\gamma^2_{12}(\bar{t})}{\gamma^2_{22}(\bar{t})}{\gamma^2_0}
}{|{G_{\rm C}}^{(0)r}(\varepsilon_f,\bar{t})|^2_{5 \times 5}}.
\end{equation}
This equation explains the response of the device in terms of the time-dependent hopping integrals
and the on-site energy.

\subsection{The first order contribution to the current}

The first order contribution to the current is expressed as

\begin{equation}
\fl
I^{(1)}(\bar t )=\frac{ie}{\hbar}\int\frac{d\varepsilon}{2\pi}{\rm Tr}\bigg\{\frac{\Gamma_{\rm L}(\varepsilon)\Gamma_{\rm R}
(\varepsilon)}{\Gamma_{\rm L}(\varepsilon)+\Gamma_{\rm R}(\varepsilon)}\bigg({G}^{(1)r}_{\rm C}(\varepsilon,
\bar t)-{G}^{(1)a}_{\rm C}(\varepsilon, \bar t)\bigg)\bigg\}[f_{\rm L}(\varepsilon)-f_{\rm R}(\varepsilon)].
\end{equation}

The first order current $I^{(1)}(\bar t)$ is found small as compared to the zeroth order current
$I^{(0)}(\bar{t})$. The first order Green's function $G^{(1)}(\bar t)$
gives rise to the first order current $I^{(1)}(\bar t)$, and it is defined as $G^{(1)r,a}(t-t^{\prime},\bar{t})=(\frac{t^{\prime}-t}{2})
\frac{\partial G^{r,a}}{\partial\bar{t}}(t-t^\prime,\bar{t})|_{\bar{t}=t}$ where the fast time variable $(t-t^\prime)$
is of the order of $\sim 10^{-15}$ s. Therefore, the first order contribution being very small is not included in the
final expression for the total current.
There could be future experiments in which the first order term is enhanced (\emph{e.g.,} when $\frac{\partial
G^{r,a}}{\partial \bar t}\sim \frac{1}{t^\prime-t}$) and the first order contribution becomes significant.

\subsection{Theory and Experiment}

The theory calculates the normalized conductance ($\cal G$$^{(0)}(\bar{t})$/$\cal G$$_0)$ explicitly for
a 5 $\times$ 5 matrix using Eqs. (59) and (60) for an $\rm Au_{20}$ cluster using the adiabatic expansion
which integrates out the fast variables of the electrons and gives the conductance in terms of the slow
time scale of seconds of the gas flow. The experimental result (Fig. 4(c) of Ref. [10]) is for an ensemble
of many SWCNTs and Au clusters having different dimensions with 30 minutes of exposure to CO gas. The 
theoretically calculated averaged conductance for a generic SWCNT with an $\rm Au_{20}$
cluster at the slow time scale is taken over many seconds to give the experimental conductance in minutes.
The term $\cal G$$_0$ is the conductance for the bare system when exposed to $\rm N_2$ and $T(\varepsilon_f,\bar t)$
is the transmission function of the Au-decorated SWCNT device when exposed to CO gas which is calculated at
different times $\bar t$ with different Fermi level $\varepsilon_f$ of the SWCNT. $T_0(\varepsilon_f^\prime,
\bar t_{\rm max})=\frac{4~{\rm Im}{\Sigma^r_{\rm L}}~{\rm Im}{\Sigma^r_{\rm R}}~{\gamma^{2}_{0}}{\gamma^{2}_{0}}{\gamma^{2}_{0}}{\gamma^{2}_{0}}}{|{G_{\rm C}}^{(0)r}(\varepsilon_f^\prime,
\bar t_{\rm max})|^2_{5 \times 5}}$ is the transmission function of the bare device when exposed to $\rm N_2$
and calculated at a maximum time $\bar t_{\rm max}$ when all $\rm N_2$ molecules interact with the $\rm
Au_{20}$-SWCNT system with a fixed value of the Fermi level $\varepsilon_f^\prime$.

As we are interested in electronic transport properties (conductance), the effect of the hopping
integral, i.e., the hopping of electrons between nearest-neighbors is considered more significant
than the on-site energy. Hence, the contribution of the on-site energy to the conductance is
suppressed. To analyze the sensor response in terms of the hopping integral ($\gamma_{\rm ij}$),
a form for $\gamma_{\rm ij}$ is considered. The hopping integral is defined as
$\gamma=<\varphi_{\rm A(r-R_A)}|H_{\rm C}|\varphi_{\rm B(r-R_B)}>$, where $\varphi_{\rm A}$
and $\varphi_{\rm B}$ are the atomic wavefunctions of the carbon atoms A and B, and
$R_{\rm A(B)}$ are their position vectors. Using the standard wavefunctions, $\gamma$
can be found to have the form $\gamma_{\rm ij}(\Delta {\rm a}(\bar{t}))= \gamma_0 \rm
exp(-\Delta a(\bar{t})/{\rm a_0})$, where the parameter $\gamma_0=2.0$ eV is the hopping
integral without the Au and CO molecules and $\rm a_0=0.33 \rm \AA$ \cite{34}.
$\gamma_{\rm ij}(\Delta \rm a(\bar{t}))$ is the modified hopping integral due to the
interaction of CO molecules with the Au-SWCNT system at each $\bar{t}$. Referring to
Fig. 2, 
when CO gas interacts with the cluster at a particular instant of time (say $\bar t_2$),
the change in the first nearest-neighbor C-Au distance $\rm \Delta a_{B_1A2}(\bar t_2)$
and $\rm \Delta a_{A_2B_2}(\bar t_2)$ are considered to be equivalent to $\rm \Delta a(\bar t_2)$
which is a parameter that includes the effects of the CO and Au on the charge distribution
and the nearest-neighbor C-Au distance. Here $\rm \Delta a_{B_1A_2}(\bar t_2)$=$\rm \Delta
a_{A_2B_2}(\bar t_2)$ as the same gas CO interacts with the same $\rm Au_{20}$ cluster which
causes same changes in the C-Au distance. But the change in the C-Au distance ($\rm a_{cAu}$)
is different at different times because as the time changes more CO molecules interact with
the Au-decorated SWCNT. In the experiment \cite{10}, the calculations show that adsorption at the
$\rm {Au_{20}}$ corner sites is the most energetically favorable configuration for the adsorption
of CO molecules than the edge sites, however, the experimental result is for many larger Au
clusters. At any instant of time, the CO molecules can be on a combination of corner and edge sites
of the Au clusters hence the different changes in the conductance arising due to the CO molecules
at the edge and corner sites is neglected and is assumed to be an average value.

In the calculation, the parameters $\Delta \rm a(\bar{t})$ and $\varepsilon_f(\bar{t})$
are found for each interaction time. The best fit value of the Fermi level ($\varepsilon_f^\prime$)
when the bare device is exposed to $\rm N_2$ is found to be -0.4 eV. 
A set of the values of $\Delta \rm a(\bar{t})$ and $\varepsilon_f(\bar{t})$ is chosen which
best fits the experimental result. These parameters carry the information about the dopants,
Au and CO molecules. The effect of these molecules on the conductance of the SWCNT can be seen
from the variation of these parameters with time, Fig. 3, causing a decrease in the conductance.
The interaction of CO molecules with the Au-decorated SWCNT at each $\bar t$ causes charge
redistribution in the system, leading to a partial charge transfer from the Au-CO complex to
the SWCNT. This deforms SWCNT and changes the nearest-neighbor C-Au distance and affects the
wavefunctions and the hopping integral, and hence changes the sensor response. This charge
transfer and the resulting changes in C-Au distance increase with increasing time as more and
more CO molecules interact with the $\rm Au_{20}$-SWCNT, as indicated by increasing negative values of
$\Delta \rm a(\bar{t})$ with time $\bar t$ in Fig. 3(a). This enhances the hopping integral
$\gamma_{\rm ij}(\Delta \rm a(\bar t)) = \gamma_0\rm exp (\Delta a(\bar t)/\rm a_0)$ as shown
in Fig. 3(b), which represents the variation of the hopping integral with respect to the CO
exposure time and the inset shows its variation as a function of the parameter $\Delta \rm
a(\bar t)$. As a result, the energy bands of the central SWCNT change locally which shifts the
position of the Fermi level away from the valence band of the SWCNT indicated by increasing values
of $\varepsilon_f$ with $\bar t$ in Fig. 3(c) and with $\Delta \rm a(\bar t)$ in the inset. This
reduces the hole carrier concentration and hence the conductance of the SWCNT, which is consistent
with the experiment. Figure 4 compares the plot between the normalized electrical conductance,
calculated by fixing  $\Delta \rm a(\bar{t})$(\AA) and $\varepsilon_f(\bar{t})$(eV), and the
experimental conductance with respect to exposure time of CO gas. Hence, the formula reproduces
the normalized electrical conductance plot of the experiment.

\begin{figure}[h!]
\center
\includegraphics[width=7cm]{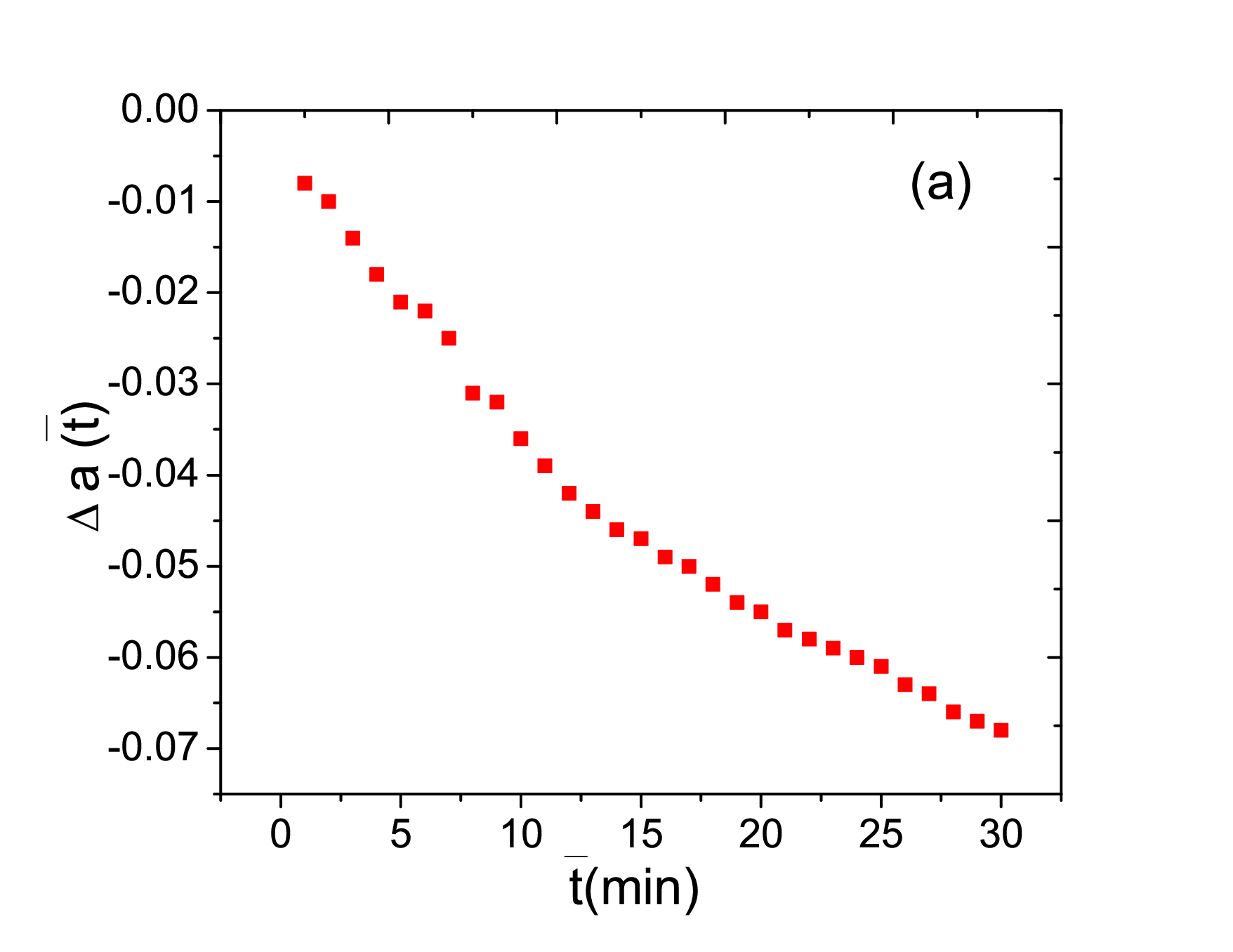}
\includegraphics[width=7cm]{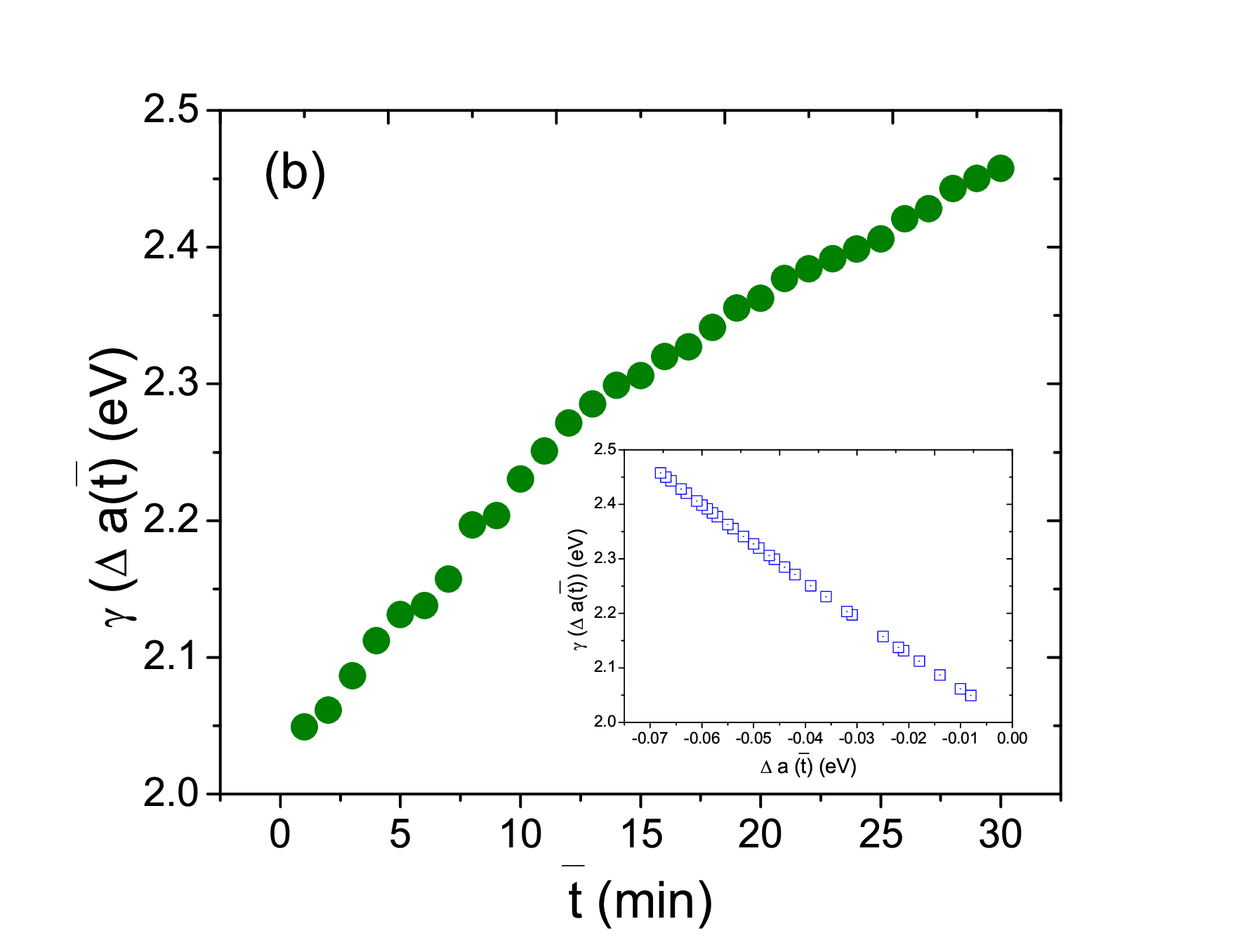}
\includegraphics[width=6.2cm]{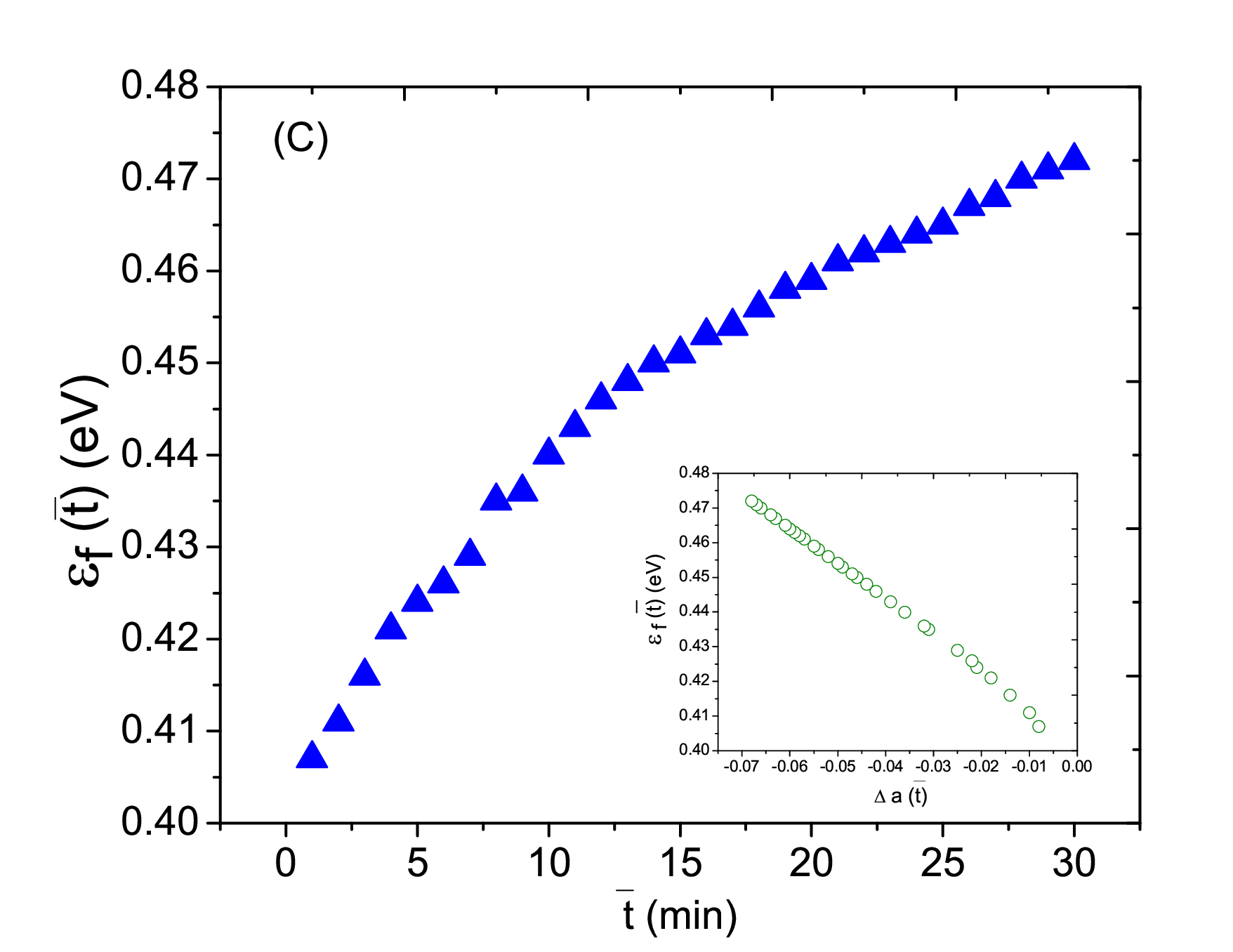}
\caption{(Color online) Variation of (a) $\Delta \rm a(\bar t)$(\AA) with time $\bar t$(min), (b)
$\gamma_{\rm ij}(\Delta \rm a(\bar t))$(eV) with time $\bar t$(min) and $\Delta \rm a(\bar t)$ (inset)
and (c) $\varepsilon_f(\bar t)$(eV) with time $\bar t$(min) and $\Delta \rm a(\bar t)$(\AA) (inset).}
\end{figure}

\begin{figure}[h!]
\center
\includegraphics[width=7.5cm]{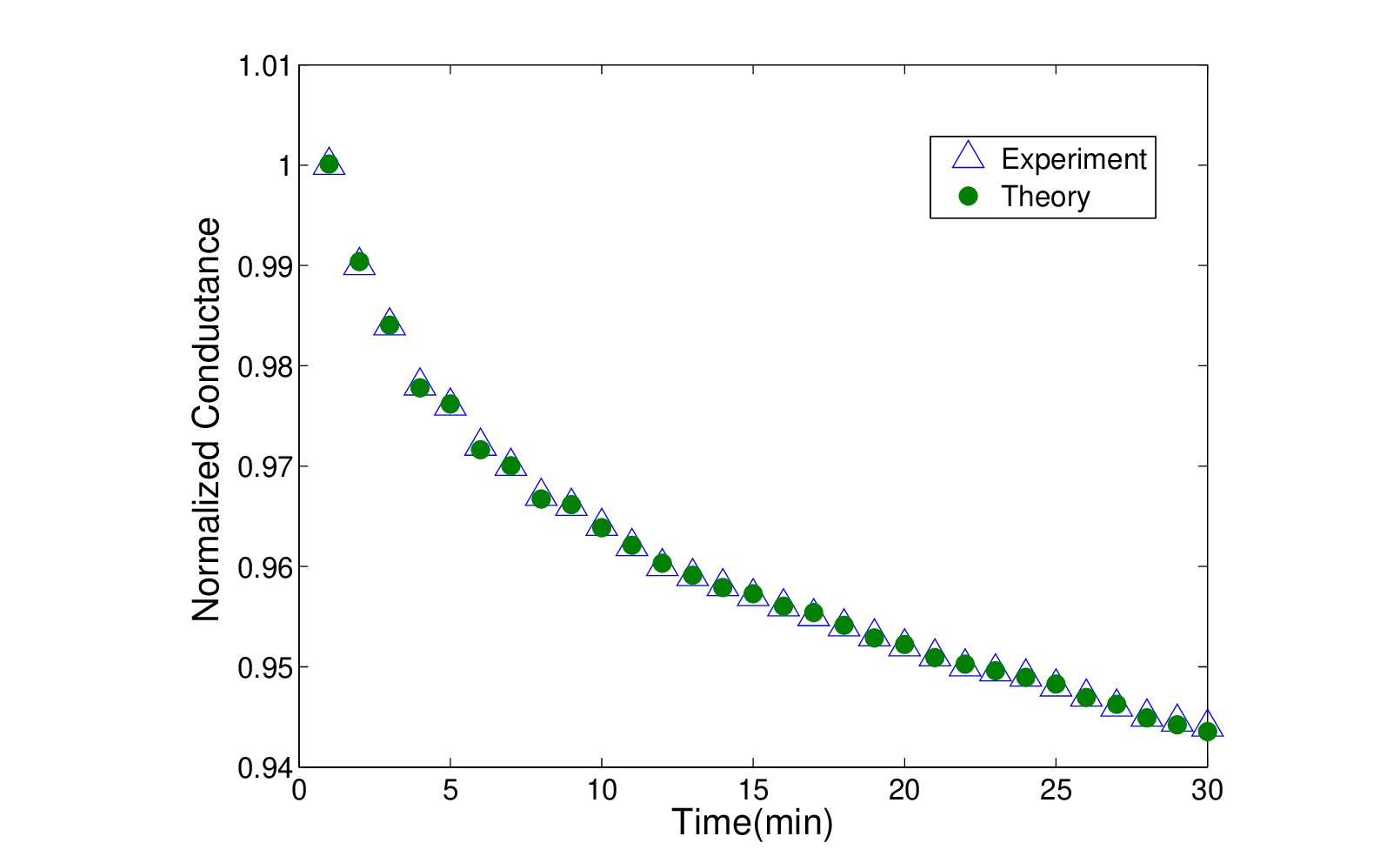}
\caption{(Color online) The average sensor response of Au-decorated SWCNTs with CO molecules adsorbed at the Au surface.
The experimental data are reproduced from Ref. \cite{10} with permission from American Chemical Society.}
\end{figure}

\subsection{Autocorrelation function}

We use a numerical and graphical technique to study the sensor response by calculating
the fluctuations, autocorrelation function (ACF) \cite{35}, of the sensor response.
The autocorrelation function is defined as
\begin{equation}
{\rm ACF}=\frac{\sum_{\bar t=1}^{{\rm n}-{\rm k}}({\cal G}(\bar t)-\bar {\cal G})({\cal G}({\bar t}+{\rm k})-\bar{\cal G})}{\sum_{\bar t=1}^{{\rm n}}({\cal G}(\bar t)-\bar{\cal G})^2},
\end{equation}
where 
${\cal G}(\bar{t})={\cal G}^{(0)}(\bar{t})/{\cal G}_0$ is the normalized conductance at
time $\bar t$ and $\bar{\cal G}={1\over {\rm n}}\sum_{\bar t=1}^{{\rm n}} {\cal G}(\bar t)$
is the mean of the conductance, n is the total number of data points and ${\rm k}=1,2,3 \cdots
{\rm K}$ is the time lag, where K $<$ n.

The autocorrelation is a mathematical tool which is used for finding patterns and degree of
randomness in a series of values (\emph{e.g.,} time series). It represents the correlation
of a variable with itself at two different times \cite{35}. Figure 5 shows the ACF plot which
starts with a positive autocorrelation that decreases gradually and becomes negative as the
time lag increases. The ACF plot gives a distinct pattern for the conductance which has a
signature between moderate and strong autocorrelation function \cite{36}, indicating non-randomness
in the sensor response and signifies the sensitivity of the sensor. Hence, ACF is useful in
finding the pattern of the sensor response. Again we find that theory reproduces experiment
in the ACF plot. This result suggests that ACF (current fluctuations) along with the current
characteristics are useful to study the sensitivity of the device by comparing the pattern of
the sensor response for different gases and may also be useful for other low dimensional systems
such as sensors based on graphene \cite{37}.

\begin{figure}[h!]
\center
\includegraphics[width=8cm]{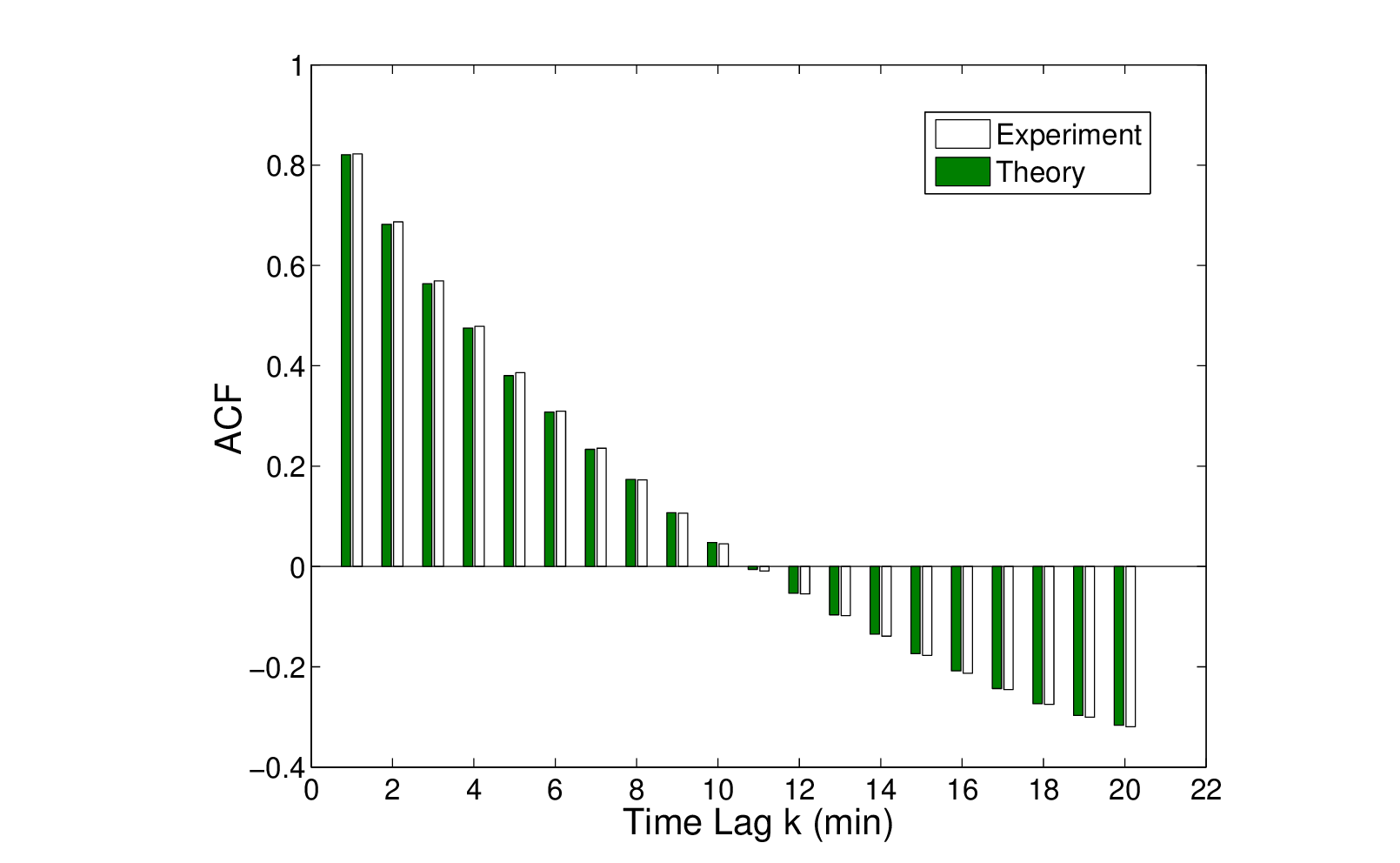}
\caption{(Color online) Experimental (empty bars) and theoretical (solid bars) autocorrelation function plots.}
\end{figure}

\section{Conclusions}
We presented a theoretical understanding of the electronic transport in the Au-decorated
SWCNT during CO gas adsorption using the NEGF formalism. To model the electronic transport
in the $\rm {Au_{20}}$-SWCNT a tight-binding model of the system was presented which
consisted of only a few nearest-neighbor carbon atoms of which only the the first nearest-neighbor
carbon atoms was affected by the $\rm Au_{20}$ cluster. The second nearest-neighbor carbon atoms
remained unaffected. For this model, we assumed that when a carbon atom of the SWCNT was
replaced by the Au cluster then the interaction of the CO gas molecules with the Au-SWCNT
system changed the hopping integral and the on-site energy as a function of interaction time
which led to a change in the conductance. In the construction of the tight-binding Hamiltonian
for the SWCNT, the effect of the interaction of CO gas with the Au-SWCNT system was
represented by taking a time-dependent Hamiltonian for the SWCNT, $H_{\rm C}$(t), where changes
due to dopants were incorporated in the changes in the hopping integrals and the on-site energy
as a function of time. 
This was because, as the time changed, the number of CO molecules and hence their interaction
with the Au cluster attached to the SWCNT changed, which resulted in a change in the on-site
energy and the hopping integral leading to a change in sensor response (conductance). But the
contact and tunneling Hamiltonian were considered to be time-independent in the absence of any
external time-dependent bias applied between the left and right contacts, and a gate potential.
Since in this problem the experimental time scale of the CO interaction with the Au functionalized
SWCNT was much longer than the time scale of electron transport inside the nanotube, therefore the
technique of the adiabatic approximation was applied to study the electronic transport
in such a gas sensor.
More general approaches to study nonequilibrium systems are given in the literature and
include the projective method \cite{38,39} with finite memory effects. The results for
transport (without memory effects) derived here by the Keldysh formalism are equivalent
to the projective approaches as large separation of time scales are exhibited in the sensors.
To calculate the conductance explicitly, we derived an explicit formula
for the transmission function in terms of the time-dependent hopping integrals and the on-site energy
in the linear response regime as the experiment was performed with low bias. The wide-band approximation
was considered in which we neglected the real part ($\Lambda$) of the self-energy and considered
the imaginary part ($\Gamma$) to be energy independent in order to obtain explicit analytic results.

To calculate the electronic transport, conductance, we assumed that the effect of the hopping
of electrons between nearest-neighbors was more significant than the on-site energy. Hence, the
contribution of the on-site energy to the conductance was suppressed. We calculated the conductance
by considering a form for the hopping integral which included the effect of the Au and CO molecules
on the charge distribution of the carbon atoms and the nearest-neighbor C-Au distance. We also assumed
that at a particular instant of time when CO gas interacted with the cluster the change in the first
nearest-neighbor C-Au distances were considered to be equivalent as the same gas CO interacted with
the same $\rm Au_{20}$ cluster which caused the same changes in the C-Au distance. But the change in
the C-Au distance ($\rm a_{cAu}$) was different at different times because as the time changed more
CO molecules interacted with the Au-decorated SWCNT. The experiment \cite{10} was done for many larger
Au clusters and the calculations showed that adsorption at the $\rm {Au_{20}}$ corner sites was the
most energetically favorable configuration for the adsorption of CO molecules than the edge sites.
Hence, at any instant of time, the CO molecules could be on a combination of corner and edge sites of
the Au clusters. Therefore, the different changes in the conductance arising due to the CO molecules at
the edge and corner sites were neglected and assumed to be an average value.
In the calculation, 
we calculated the normalized conductance explicitly for an $\rm Au_{20}$ cluster. The experimental
result of the conductance in the time scale of minutes was produced by the theoretically calculated averaged
conductance for a generic SWCNT with an $\rm Au_{20}$ cluster at the slow time scale taken over many seconds.
The parameters $\rm \Delta \rm a(\bar{t})$ and $\varepsilon_f(\bar{t})$ were found for each interaction 
time to match the experimental data very well. The predicted values of these
parameters gave the electronic band structure and hence the electronic transport properties of the SWCNT
when CO molecules interacted with the Au functionalized SWCNT. The prediction, $\rm \Delta \rm a(\bar{t})$
and $\varepsilon_f(\bar{t})$, of the model can be verified by experiments, ab-initio technique or density
functional theory based calculations. More sophisticated methods should be tried on these systems for better
modeling the Au clusters and this is left for future studies.

The set of Eqs. (56), (60), and (61) with derivations of the equation of motion and Dyson
equation were the major results of this manuscript and have been reported for the Au-SWCNT system
used for CO gas detection. Equation (56) represented the zeroth order current which
is the time-dependent Landauer formula. The dependence of the transmission function and conductance
on $\gamma(\bar{t})$ and $\varepsilon_{\rm A/B}(\bar{t})$ was shown by Eq. (60).
The first order contribution to the current was also derived which was found small, Eq. (61). Hence
this phenomena was essentially described by the Landauer formula but evaluated at each time $\bar t$.
The formula for the transmission function was then used to compare the theoretical results with the
experiment. We found the formula quantitatively reproduced the experimental result. The model gave a
set of parameters that best fitted the experiment. We also calculated and compared the autocorrelation
function for the theoretical and experimental sensor response and found the results were in agreement.
We observed a correlation in the sensor response indicating the sensitivity of the Au-decorated
SWCNT sensor. Hence, the autocorrelation function was found useful in recognizing the pattern of the
sensor response.

This work is an attempt at a microscopic study of time-dependent electronic transport in Au-decorated
SWCNT sensors using the NEGF formalism and presents an important contribution towards a conceptual
understanding of how the Au and CO molecules change the experimentally measured conductance of the
nanotube and hence developed a connection between the experimental observations and theoretical
predictions. Such a detailed investigation is needed which may give insight into a microscopic
understanding of the fundamental science of molecule-CNT nanohybrids and electronic transport in such
hybrid systems for
beneficial applications, and could strengthen the experimental results. This theoretical approach can be
applied to other low dimensional systems such as graphene and CNTs functionalized with DNA \cite{37,40}
for sensing applications.


\section*{Acknowledgments}
This work has been supported by the Council of Scientific and Industrial Research (CSIR),
the University Grant Commission (UGC) and the University Faculty R $\&$ D Research Programme.

\end{document}